\documentclass[a4,aps,pra,reprint,longbibliography,floatfix,table]{revtex4-2}
\usepackage{amsmath,amsthm,amsfonts,amssymb,graphicx,hyperref,epstopdf,xcolor,tikz,scalerel,natbib,array}
\usepackage{mathptmx,textcomp}
\usepackage[T1]{fontenc}
\usepackage[utf8]{inputenc}

\newcommand{\td}{\textdegree}
\definecolor{lime}{HTML}{A6CE39}

\usetikzlibrary{svg.path}
\definecolor{orcidlogocol}{HTML}{A6CE39}
\tikzset{
orcidlogo/.pic={
\fill[orcidlogocol] svg{M256,128c0,70.7-57.3,128-128,128C57.3,256,0,198.7,0,128C0,57.3,57.3,0,128,0C198.7,0,256,57.3,256,128z};
\fill[white] svg{M86.3,186.2H70.9V79.1h15.4v48.4V186.2z}
svg{M108.9,79.1h41.6c39.6,0,57,28.3,57,53.6c0,27.5-21.5,53.6-56.8,53.6h-41.8V79.1z M124.3,172.4h24.5c34.9,0,42.9-26.5,42.9-39.7c0-21.5-13.7-39.7-43.7-39.7h-23.7V172.4z}
svg{M88.7,56.8c0,5.5-4.5,10.1-10.1,10.1c-5.6,0-10.1-4.6-10.1-10.1c0-5.6,4.5-10.1,10.1-10.1C84.2,46.7,88.7,51.3,88.7,56.8z};
}
}
\newcommand\orcidicon[1]{\href{https://orcid.org/#1}{\mbox{\scalerel*{
\begin{tikzpicture}[yscale=-1,transform shape] \pic{orcidlogo}; \end{tikzpicture}
}{|}}}}  

\DeclareMathAlphabet{\mathpzc}{OT1}{pzc}{m}{it}
\hypersetup{colorlinks,    citecolor=blue,    filecolor=black,    linkcolor=blue,    urlcolor=blue}
\begin{document}
\title{Twisted electron impact single ionization coincidence cross-sections for noble gas atoms}
\author{{Nikita Dhankhar \orcidicon{0000-0001-9423-4796}}}
\author{{Soham Banerjee \orcidicon{0000-0002-6400-3580} }}

\author{{R. Choubisa \orcidicon{0000-0003-3000-6174}} }
\email{rchoubisa@pilani.bits-pilani.ac.in}
\affiliation{ Department of Physics, Birla Institute of Technology and Science-Pilani, Pilani Campus, Pilani,  Rajasthan, 333031, India}

\begin{abstract}
We present the angular profiles of the triple differential cross-section (TDCS) for the (e,2e) process on the noble gas atoms, namely He (1s), Ne (2s and 2p), and Ar (3p) for the plane wave and the twisted electron impact. We develop the theoretical formalism in the first-born approximation.
The present study compares the TDCS for different values of orbital angular momentum number {\it m} and opening angles $\theta_p$ of the twisted electron beam with that of the plane wave beam. In addition, we also investigate the TDCS for macroscopic atomic targets to explore the influence of opening angle $\theta_p$ of the twisted electron beam on the TDCS. 
Our results show that the peaks in binary and recoil region shift from the momentum transfer direction. The results also show that for larger opening angles the peaks for {\it p}-type orbitals split into double-peak structures which are not observed in the plane wave results for the given kinematics. The angular profiles for averaged cross-section show the dependence of TDCS on the opening angles which are significantly different from the plane wave TDCS.

\end{abstract}

\maketitle

\section{Introduction}
The process of electron impact ionization (hereafter referred as an (e,2e) process) of various atomic and molecular targets frequently finds an important place in many fields. To name a few, electron impact ionization has practical applications in plasma physics, atmospheric physics, radiation physics, astronomy, and even biology \cite{Bartschat2016, DUNN2015, Girazian2017, Kyniene2019, Caleman2009}. 
A single electron is ejected from the target in a coincident (e,2e) process when interacting with an incident electron.  The ejected and scattered electrons are detected with their momenta fully resolved \cite{Camp2018}. Experimental research in this field was first performed by Erhardt {\it et al.} \cite{Ehrhardt1969} and Amaldi {\it et al.} \cite{Amaldi1969} independently. Since then, there has been tremendous progress in the theoretical and experimental study of electron impact ionization processes. The way we generally study an (e,2e) process is through the triple differential cross-section (TDCS). The TDCS is proportional to the probability of detecting the two outgoing electrons (the ejected electron and the scattered electron) in coincidence with their momenta fully resolved. Measuring and analyzing the TDCS of (e,2e) processes can provide valuable insight into collision mechanisms and electronic structure of the given target and serve as very strict tests on theoretical models of few-body quantum dynamics.\\

After decades of progress in theoretical models aimed at describing (e,2e) processes, we now have models that can fully explain the cross-sections of simple targets such as H and He atoms \cite{Madison1984, CCC2009}. However, current theoretical models still struggle to fully understand the dynamics for more complex targets, such as molecules and multi-electron atoms. Among these complex targets are inert gas atoms, such as Ne, Ar and Xe. Theoretical models for atomic H and He that give good agreement with experimental results, such as the BBK method \cite{BBK1999}, yield discrepancies for heavier atoms in the intermediate and low energy regimes. Significant efforts have been made to develop theoretical models that agree with experimental TDCS results of inert atoms. Bell \emph{et al.} \cite{Bell1995} measured experimental TDCS results for Argon (3p) electrons in co-planar geometry and compared them with theoretical results using distorted wave Born approximation (DWBA). Brothers \emph{et al.} \cite{MJ1986} calculated theoretical cross sections for Ne (2s), Ne (2p), Ar (2p), and Ar (3p) using the factorized plane-wave Born approximation. The BBK and the DWBA$+R$-matrix (close coupling) models were also used to describe TDCS of He (1s) and Ar (3p) \cite{Catoire2006}. Kheifets \emph{et al.} \cite{Kheifets2008} used the DWBA-G (DWBA corrected by the Gamow factor) model to include the post-collision interactions with the Gamow factor for calculating TDCS of He (1s), Ne (2s), Ne (2p), and Ar (3p). Other sophisticated models, like a convergent close-coupling, DWBA-G, and a hybrid DWBA$+R$-matrix method, are also used for the calculation of the TDCS for Ne and Xe ionization process \cite{Stevenson2009}. Pflüger \emph{et al.} \cite{Pfluger2013} used the non-perturbative B-spline R Matrix (BSR \cite{BSR2013}) approach  to calculate low incident energy Ne (2p) TDCS. Ren \emph{et al.} \cite{Ren2015} used different models to study Ne (2p) ionization at low energies, including second order DWBA-RM model, BSR model, three-body distorted wave (3DW) model and distorted-wave Born approximation with Ward-Macek (DWBA-WM) approximation. 
The TDCSs have also been calculated for Ar (3p) and Ar (3s) atoms  in the modified distorted wave formalism using the second Born approximation \cite{Purohit2020}.
Very recently, Gong \emph{et al.} \cite{Gong2020} used the multi-center three distorted wave method(MCTDW) to calculate Ne(2p) TDCS at low incident energies.\\

There have been several undertakings to improve the agreement between experimental and theoretical data for the plane wave ionization of inert atoms. However, it is still to be explored for the twisted beam impact ionization of the same. Unlike plane waves, twisted electron beams carry a well-defined orbital angular momentum (OAM) along the beam axis. This non-zero OAM stems from the usual phase factor of $e^{im\phi}$. In addition to quantized OAM, they also have a non-zero transverse linear momentum, and non-uniform intensity in the transverse direction \cite{Bliokh2017,Harris2019}. Inspired by optical vortex beams, many experimental groups have recently demonstrated the production of electron vortex beams with several orders of phase singularity and the corresponding angular momentum\cite{Uchida2010, Verbeeck2010}. Different experimental groups have been successful in generating electron vortex with very high angular momentum: up to $1000\hbar$ \cite{Benjamin2011,Mafakheri2017}. These laboratory realizations of vortex beams have stirred up a brand new area of research and boast promising applications. There is a scope for applications in transmission electron microscopy(TEM), fine probing of matter, surveying magnetic properties of novel materials, investigating chiral properties of crystals, strong-field ionization, transfer of angular momentum to nanoparticles, etc.\cite{Boxem2016, Asenjo2014}.
The quantized OAM possessed by electron vortex beams acts as a new degree of freedom in electron-target collision studies. Changing this new degree of freedom may have interesting effects, and therefore, we must investigate the impact of OAM on collision cross-sections.

 Pioneering studies by Boxem {\it et al.} \cite{Boxem2014,Boxem2015,Boxem2016} inspired further research in this direction.
Serbo {\it et al.} \cite{Serbo2015} analyzed twisted electron scattering in the relativistic regime and showed that the cross-section for bulk targets must be integrated over all possible impact parameters. This integral eliminates any dependence on the `{\it m}' value or the amount of OAM associated with the twisted beam. Matula {\it et al.} \cite{Matula2014} studied the role of angular momentum in radiative capture of twisted electrons by bare hydrogen-like ions. Using a theoretical analysis of twisted Electron Energy-Loss Spectroscopy (EELS) from C$_{60}$ fullerene, Schüler and Berakdar investigated the effects of OAM transfer on plasmon generation \cite{Schuler2016}. Koshleva {\it et al.} calculated the cross-sections for twisted electron Mott scattering by atomic targets in the fully relativistic regime \cite{Kosheleva2018}. Using the first Born approximation, Maiorova {\it et al.} investigated the elastic scattering of twisted electrons by $H_2$ molecule \cite{Maiorova2018}. Harris {\it et al.} \cite{Harris2019} calculated the fully differential cross-section corresponding to twisted electron wave ionization of hydrogen atom. The FDCS showed a shift in recoil and binary peak positions due to the transverse momentum component.  Dhankhar \emph{et al.} reported theoretical calculations of the five-fold differential cross-section of double ionization of He atom by twisted waves in $\theta$ variable mode and constant $\theta_{12}$-mode \cite{Dhankhar2020}.  The study by Mandal \textit{et al.} (2020) showed the dependence of the Total Angular Momentum (TAM) number (\textit{m}) and opening angle ($\theta_p$) on the angular profile of the TDCS and spin asymmetry for the relativistic electron impact ionization of the heavy atomic targets \cite{Mandal2020}. Dhankhar \emph{et al.} investigated the influence of the OAM number `{\it m}' and the opening angle $\theta_p$ single ionization cross-sections of $H_2$ and $H_2O$ molecule by twisted electron waves independently \cite{Dhankhar2020_2,Dhankhar2021}. 

In this communication, we extend our previous studies (for (e,2e) process on  $H_2$ and $H_2O$) of the differential cross-section by the twisted electron beam on the inert gas atoms.
We present the theoretical estimation for the ionization of noble gas atoms, namely He, Ne (2s and 2p), and Ar (3p) for the twisted electron beam. Our theoretical calculations are within the first Born approximation framework for an incident plane wave electron beam and an incident twisted electron beam. We describe the plane wave, Roothan-Hartree-Fock wavefunction, Coulomb wave for the scattered electron, the bound states of He, Ne, and Ar, and the ejected electron, respectively. We present the theoretical formalism of our calculation of the TDCS in Sec.\ref{sec2}. We present our results of the TDCS for the different outer orbitals of the atoms for different parameters of the twisted electron beam in Sec.\ref{sec3}. Finally, we conclude our paper in the Sec.\ref{sec4}. Atomic units are used throughout the paper unless otherwise stated.

\section{Theoretical Formalism}\label{sec2}
In this study, we have calculated the TDCS for (e,2e) process on inert gas atoms :
\begin{equation} \label{eq1}
e_{i}^{-} + X \rightarrow e_{s}^{-} + e_{e}^{-} + X^{+}.
\end{equation}
here, $e_{i}^{-}$, $e_{s}^{-}$, $e_{e}^{-}$, {\it X} and  $X^+$ represents the incident, scattered, the ejected electron, the target atom and the ionized target respectively.
The theory used here closely follows the approach used in our previous studies \cite{Dhankhar2020_2,Dhankhar2021}.  We have carried out the theoretical analysis in the framework of the first Born approximation. We have also neglected the exchange effects between the incident/scattered electron with the bound/ejected electron here since the incident/scattered electron is faster than the bound/ejected electron ($E_s>>E_e$).\\

The triple differential cross section for an (e,2e) process of an inert gas can be written as,
\begin{equation} \label{eq2}
\frac{d^{3}\sigma}{d\Omega_{e}d\Omega_{s}dE_{e}} = (2\pi)^{4}\frac{k_{e}k_{s}}{k_i}|T^{pw}_{fi}(\mathbf{q})|^{2},    
\end{equation}

here, $d\Omega_{e}$ and $d\Omega_{s}$ denote differential intervals of the solid angles of the ejected and scattered electrons respectively. $dE_e$ denotes the energy interval of the ejected electron. $\mathbf{k}_i, \mathbf{k}_s,$ and $\mathbf{k}_e$ are the momentum vectors of the incident, scattered and ejected electrons respectively. The plane wave scattering amplitude is given by,
\begin{equation}\label{eq3}
T^{pw}_{fi}(\mathbf{q}) = \langle \phi_f|V|\phi_i \rangle,  
\end{equation}
where $\mathbf{q}=\mathbf{k}_i-\mathbf{k}_s$ is the momentum transfer vector, $\phi_i$ is the initial state wave function of the system, and $\phi_f$ is the final state wave function.
$V$ is the interaction potential between the incident electron and the multi-electron target atom. In this communication we use the frozen-core approximation, in which the problem of N (total number of electrons in the target atom) electrons can be reduced to one active electron problem. In this approximation, we assume that only one of the target electrons (from the outer orbital of the target atom) participate in the ionization process and is ejected in the final channel, while the other electrons remain as frozen. The interaction potential under the approximation is then defined as,
\begin{equation} \label{eq4}
V = \frac{-1}{r_1} + \frac{1}{r_{12}},  
\end{equation}
In the above equation, the first term represents the Coulomb interaction of the incident electron with the nucleus of the inert gas atom, with $r_1$ as the distance from the nucleus to the incident electron. The second term is the Coulomb interaction between the incident electron and the participating bound electron, with $r_{12}$ as the separation between the two electrons.\\
For calculating the transition matrix element, we have used the following assumptions:
\begin{itemize}
\item The incident and the scattered electrons are both treated as plane waves. 
\item The Roothaan-Hartree-Fock wave functions represent the participating bound electron wave function of the target atom. These atomic wave functions are given by a combination of Slater orbitals using Clementi-Roetti (CR) coefficients \cite{CR1974}, where Slater orbitals take the following general form \cite{Camp2018},
\begin{equation} \label{eq5}
\chi_{nlm}(\mathbf{r}) = N(n,\alpha) r^{n-1}e^{-\alpha r} Y_{lm}(\theta,\phi).    
\end{equation}
Here {\it n}, {\it l}, {\it m} are the quantum numbers of the atomic orbital, $\alpha$ is the orbital exponent, $Y_{lm}(\theta,\phi)$ are the complex spherical harmonics, and
\begin{equation} \label{eq6}
N(n,\alpha) = \frac{(2\alpha)^{n+\frac{1}{2}}}{[(2n)!]^{\frac{1}{2}}}    
\end{equation}
is the normalization constant.
\item The ejected electron is described by a Coulomb wave, $\Psi_{k_e}^{-}(\mathbf{r}_2)$. This way, the long-range effects of the residual ion on the slow electron are taken into account.
\end{itemize}

We can then write the initial wave function as;
\begin{equation} \label{eq7}
\phi_i = \chi_{\mathbf{k}_i}(\mathbf{r}_1)\Phi_j(\mathbf{r}_2),  
\end{equation}
where $\chi_{\mathbf{k}_i}(\mathbf{r}_1)$ is the normalized incident electron plane wave, and $\Phi_j(\mathbf{r}_2)$ is the $j^{th}$ Hartree-Fock atomic orbital from CR coefficients of the target atom and is described as;
\begin{equation} \label{eq8}
\Phi_j(\mathbf{r}_2) = \sum_{i=1}^N c_i \chi_{n_il_im_i}(\mathbf{r}_2),  
\end{equation}
where {\it i} is the $i^{th}$ basis function, $c_i$ is the expansion coefficient of the atomic orbital for the $i^{th}$ basis function and $\chi_{n_il_im_i}(\mathbf{r}_2)$ is given by equation \ref{eq5}.
Similarly, the final state wave function can be written as 
\begin{equation}\label{eq9}
\phi_f = \chi_{\mathbf{k}_s}(\mathbf{r}_1)\psi_{\mathbf{k}_e}^-(\mathbf{r}_2), 
\end{equation}
where $\chi_{\mathbf{k}_s}(\mathbf{r}_1)$ is approximated by plane wave, and $\psi_{\mathbf{k}_e}^-(\mathbf{r}_2)$ is approximated by the Coulomb wave, described as;
\begin{equation}\label{eq10}
\psi_{\mathbf{k}_e}^-(\mathbf{r}_2) = \Gamma(1-i\eta)e^{-\frac{\pi\eta}{2}}\frac{e^{i\mathbf{k}_e.\mathbf{r}_2}}{(2\pi)^{3/2}}{}_{1}F_1(i\eta,1,-i(k_er_2-\mathbf{k}_e.\mathbf{r}_2)).  
\end{equation}
Here, $\Gamma$ is the gamma function, $\eta = \frac{-Z}{k_e}$ is the Sommerfeld parameter ( Z = Nuclear charge = 1 in our case because of the frozen core approximation), and ${}_{1}F_{1}$ is the confluent hypergeometric function of the first kind.\\
The transition matrix element can be simplified by analytically performing the integral over incident electron coordinate $r_1$ \cite{Tweed1992} as follows;
\begin{equation}\label{eq11}
\int\frac{e^{i\mathbf{q.r}_1}}{|\mathbf{r}_1-\mathbf{r}_2|}d^{3}r_{1} = \frac{4\pi}{q^2}e^{i\mathbf{q.r}_2}.  
\end{equation}
The transition matrix element is then given by the following expression
\begin{equation} \label{eq12}
T^{pw}_{fi}(\mathbf{q}) = \frac{-2}{q^2} \langle \Psi^{-}_{\mathbf{k}_e}|e^{i\mathbf{q.r}_2}-1|\Phi_{j}(\mathbf{r}_2) \rangle.  
\end{equation}
The extension to twisted wave scattering in this formalism is straightforward. We change the incident electron wave function from a plane wave to a twisted wave propagating along {\it z}-axis. Similar to our previous studies, we have used the Bessel beam \cite{Boxem2015} to represent the incident electron that carries quantized angular momentum. In this case, the incident momentum vector $\mathbf{k}_i$ has a transverse component and is given by the following expression.
\begin{equation} \label{eq13}
\mathbf{k}_i = (k_i\sin{\theta_p}\cos{\phi_p})\hat{x}+(k_i\sin{\theta_p}\sin{\phi_p})\hat{y}+(k_i\cos{\theta_p})\hat{z}.  
\end{equation}
Here, $\theta_p$ and $\phi_p$ are the polar and azimuthal angles of the momentum vector respectively. $\mathbf{k}_i$ carves out the surface of a cone when we change $\phi_p$. The longitudinal component of the momentum $\mathbf{k}_{iz}$ is fixed, but the transverse momentum component $\mathbf{k}_{i\bot}$ has a direction that depends on $\phi_p$. This means that the direction of incident momentum is not well defined. However, the magnitude of the transverse momentum is fixed, and is given by $|\mathbf{k_{i\bot}}| = \varkappa = \sqrt{(k_i)^2 - (k_{iz})^2}$. The opening angle of the cone can be calculated as $\theta_p = \arctan(\frac{k_{i\bot}}{k_{iz}})$. The incident electron wave function can be written as follows;
\begin{equation} \label{eq14}
\chi^{tw}_{\varkappa m}(\mathbf{r_1}) = \int^{\infty}_{0}\frac{dk_{i\bot}}{2\pi}\int^{2\pi}_{0}\frac{d\phi_p}{2\pi}a_{\varkappa m}(k_{i\bot})e^{i\mathbf{k_i.r_1}}e^{-i\mathbf{k_i.b}}.  
\end{equation}
This can be interpreted as a superposition of plane waves with amplitude $a_{\varkappa m}(k_{i\bot})$,where this amplitude is given by;
\begin{equation} \label{eq15}
a_{\varkappa m}(k_{i\bot}) = (-i)^m e^{im\phi_p}\delta(|\mathbf{k}_{i\bot}|-\varkappa).
\end{equation}
The phase singularity of the $e^{im\phi_p}$ is responsible for the quantized angular momentum content of the twisted wave, and the delta factor ensures that the magnitude of the transverse momentum is well defined. Also, the addition of the $e^{-i\mathbf{k}_i.\mathbf{b}}$ factor accounts for the transverse orientation of the incident beam with respect to the target. Here, $\mathbf{b}$ is the impact parameter vector measured with respect to the axis of the incident wave. Unlike the plane wave, the Bessel beam has concentric rings of high and low-intensity in the transverse direction  \cite{Llyod2017}. Therefore, we need to account for the orientation of the target with respect to the incident beam. \\
\begin{figure*}
	\includegraphics[width=16cm,height=15cm]{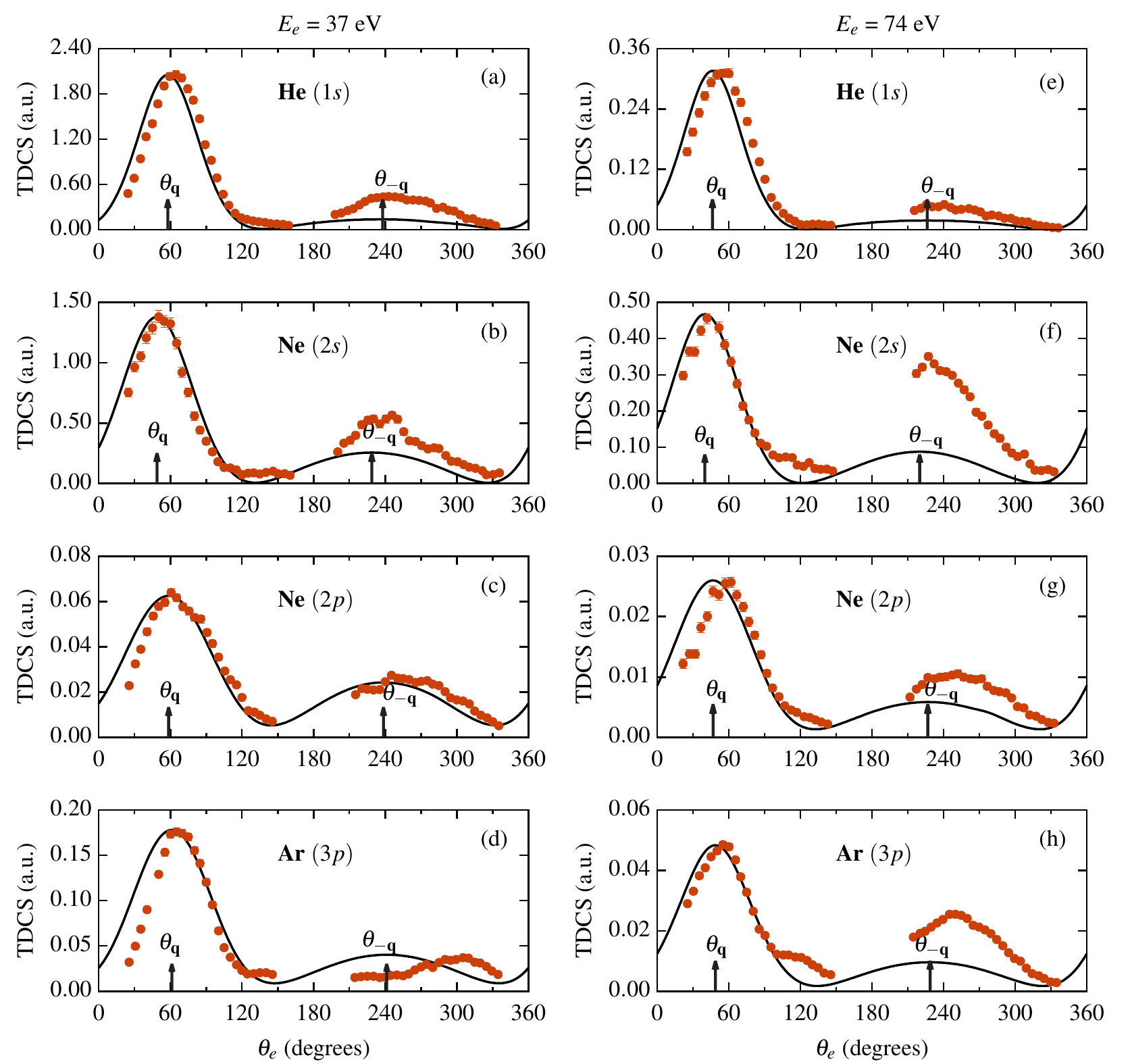}
	\caption{TDCS as a function of ejected electron angle $\theta_e$ for the plane wave (e,2e) process on He, Ne (2s and 2p shell) and Ar atoms in the co-planar asymmetric geometry. The plane wave results of our theoretical model are represented by black solid line and experimental results \cite{Kheifets2008} by full circles. The kinematics used here is : $E_s$ = 500eV, $E_e$ = 37eV and 74eV and $\theta_s$ = 6\td. The arrows indicate the direction of plane wave momentum transfer ($\theta_{\mathbf{q}}$) and recoil direction ($\theta_{-\mathbf{q}}$) for this and subsequent figures.}\label{f1}
\end{figure*}
When we replace the incident plane wave by the twisted wave, we get the following expression for twisted wave matrix element
\begin{equation} \label{eq16}
T^{tw}_{fi}(\varkappa,\mathbf{q,b}) = (-i)^m\int^{2\pi}_0\frac{d\phi_p}{2\pi}e^{im\phi_p-i\mathbf{k.b}}T^{pw}_{fi}(\mathbf{q}). 
\end{equation}

Therefore, we can evaluate the twisted wave matrix element $T^{tw}_{fi}$ in terms of the plane wave matrix element $T^{pw}_{fi}$ \cite{Dhankhar2020}. The key difference here is that the momentum transfer vector $\mathbf{q} = \mathbf{k}_i-\mathbf{k}_s$ has to be calculated using the new twisted wave momentum vector $\mathbf{k}_i$ from equation \ref{eq13}. We can then write
\begin{equation} \label{eq17}
q^2 = k^2_i + k^2_s -2k_ik_s\cos(\theta),  
\end{equation}
where,
\begin{equation} \label{eq18}
\cos(\theta) = \cos(\theta_p)\cos(\theta_s) + \sin(\theta_p)\sin(\theta_s)\cos(\phi_p-\phi_s).
\end{equation}

$\theta_s$ and $\phi_s$ are the polar and azimuth angles of the scattered electron momentum vector $\mathbf{k}_s$, which is described by a plane wave in our treatment. Unlike the plane wave, the momentum transfer of a twisted wave is not constant for a particular direction of $\mathbf{k}_s$, and depends on the azimuthal angle of the incident wave vector $\mathbf{k}_i$. This inherent uncertainty of momentum transfer direction for a twisted wave is accounted for by taking an integral over the azimuthal angle $\phi_p$ in equation \ref{eq16}.\\

We can now calculate the TDCS for twisted waves using equation \ref{eq16} and equation \ref{eq12} for the target fixed with respect to the incident beam. However, since scattering experiments are usually characterized by a uniform distribution of scattering centers ( inert atoms in our case) over a certain volume, we cannot assume a fixed value of impact parameter $\mathbf{b}$. Instead, we need to average over all possible values of $\mathbf{b}$ for meaningful results that can be compared with experimental results. The details of averaging over all possible orientations can be found in \cite{Serbo2015, Harris2019}, and average TDCS can be written as ;
\begin{equation} \label{eq19}
(TDCS)_{av}=\frac{1}{2\pi\cos\theta_p}\int^{2\pi}_{0}d\phi_p \frac{d^3\sigma(\mathbf{q})}{d\Omega_{e}d\Omega_{s}dE_{e}}.  
\end{equation}
\begin{figure*}
	\includegraphics[width=15cm,height=12cm]{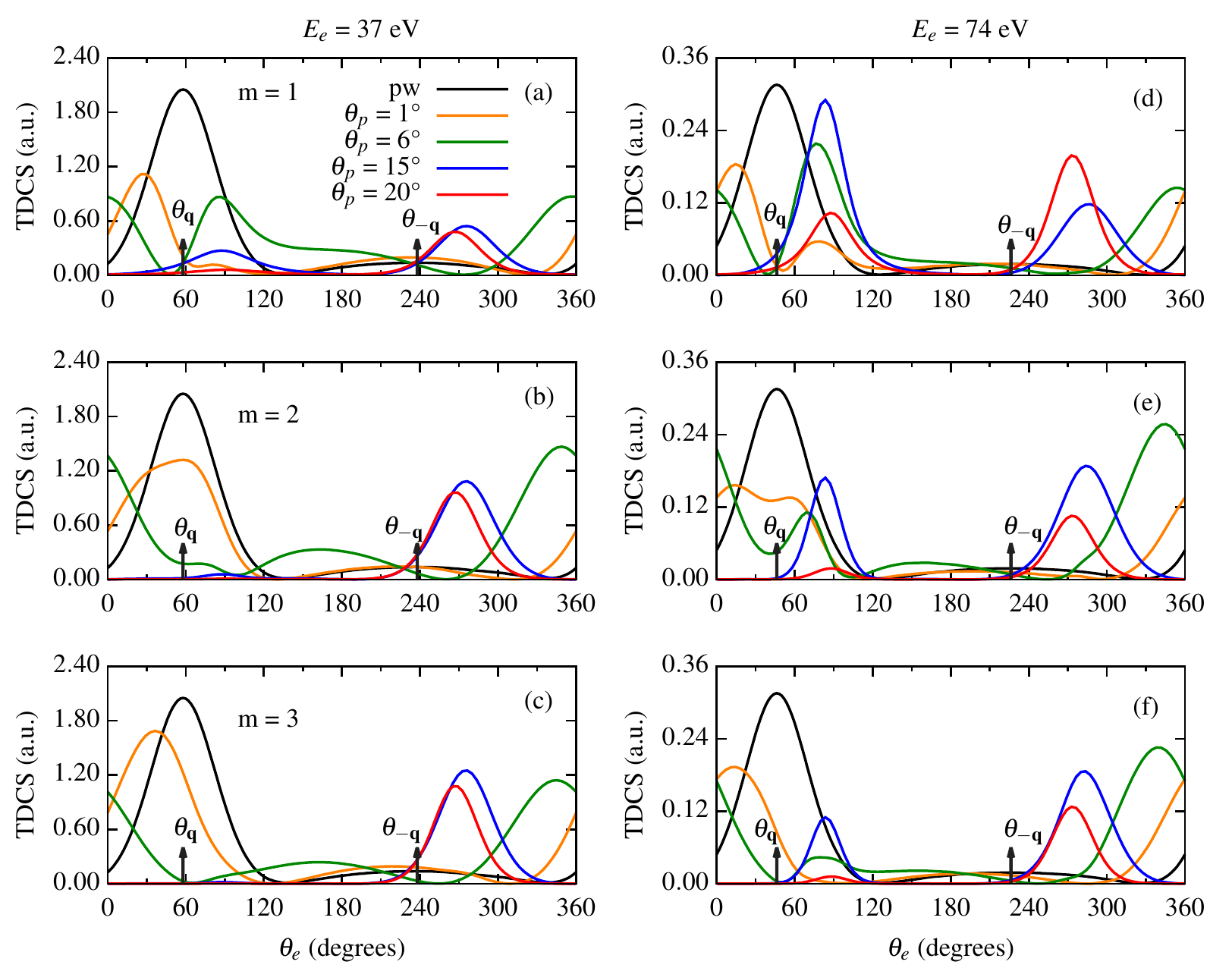}
	\caption{Angular profile of the TDCS as  a function of the ejection angle $\theta_e$ for the twisted electron wave (e,2e) process on He atom in the co-planar asymmetric geometry. The kinematics is $E_s$ = 500eV, $E_e$ = 37eV (left column) and 74eV (right column) and $\theta_s$ = 6\td. The black, orange, green, blue and red curves represent the TDCS for plane wave, $\theta_p$ = 1\td, 6\td, 15\td \ and  20\td. Sub-figures (a) and (d) are for {\it m} = 1, (b) and (e) for {\it m} = 2 and (c) and (f) are for {\it m} = 3.}\label{f2}
\end{figure*}

Here, $\frac{d^3\sigma(\mathbf{q})}{d\Omega_{e}d\Omega_{s}dE_{e}}$ is the TDCS for a particular value of $\mathbf{q}$. Interestingly, averaging over impact parameter $\mathbf{b}$ makes the TDCS independent of the OAM value `m'\cite{Serbo2015}. However, this does not mean that twisted wave TDCS is the same as plane wave TDCS since the transverse momentum component of twisted waves plays a direct role in the momentum transfer. The cross-section now depends on the opening angle $\theta_p$ of the incident twisted electron beam.


\section{Results and discussions}\label{sec3}

\begin{figure*}
	\includegraphics[width=15cm,height=12cm]{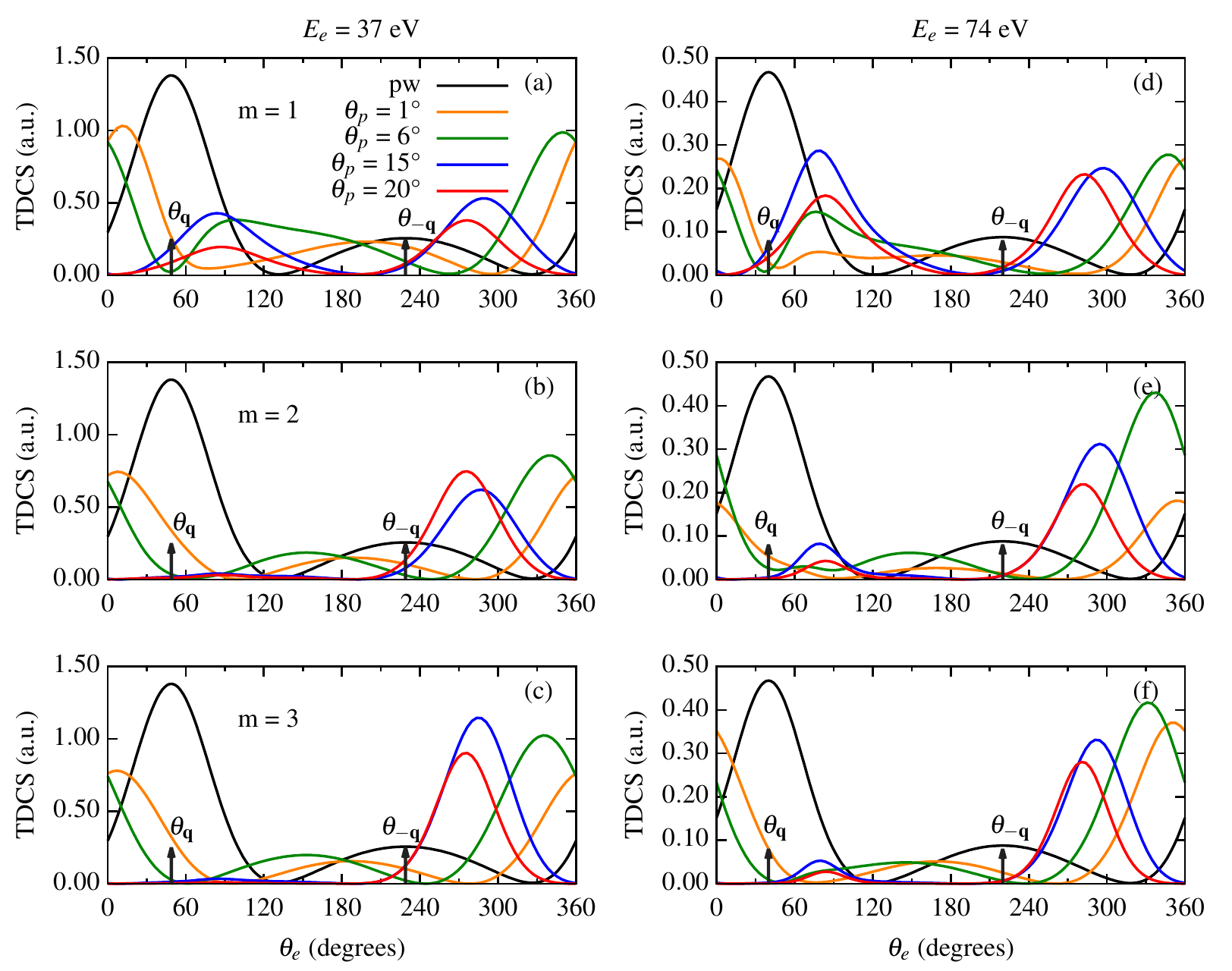}
	\caption{Same as figure \ref{f2} except that the TDCSs are plotted for (e,2e) on Ne (2s) atom.}\label{f3}
\end{figure*} 

We present the results of our calculations for the TDCS for noble gas atoms {\it viz.}, Helium (1s), Neon (2s and 2p orbitals), and Argon (3p) by a twisted electron beam impact in this section. We benchmark our theoretical calculations with the existing experimental data for the plane wave. We present the results of the TDCS for twisted electron beam to study the effects of different parameters of the beam on TDCS.
\begin{figure*}
	\includegraphics[width=15cm,height=12cm]{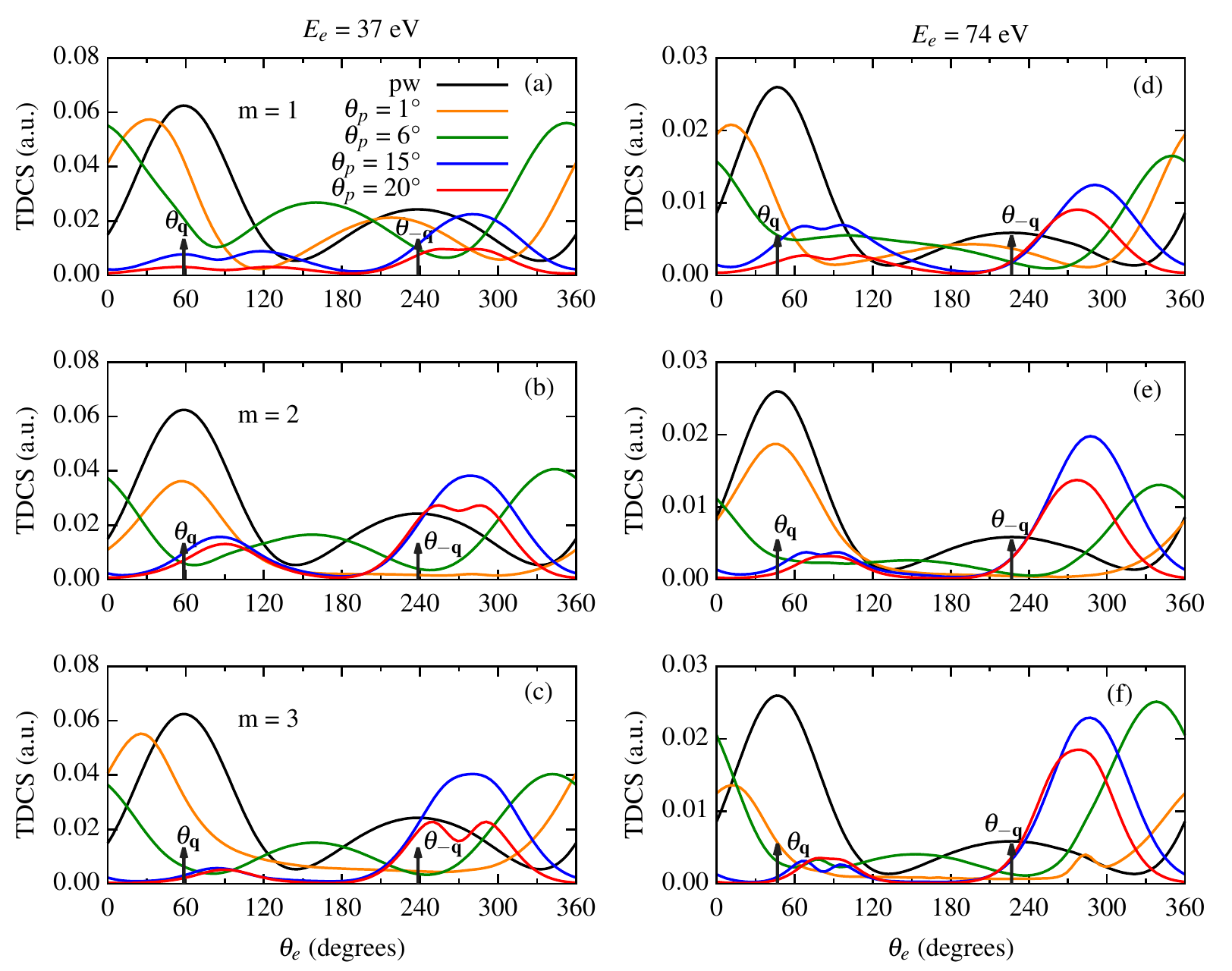}
	\caption{Same as figure \ref{f2} except that the TDCSs are plotted for (e,2e) on Ne (2p) atom.}\label{f4}
\end{figure*}
We compare our twisted electron beam TDCS with that of the plane wave TDCS for different values of orbital angular momentum (OAM) number {\it m} and opening angle $\theta_p$. For the given {\it m}, we compare the TDCS's profiles for the different values of $\theta_p$ , {\it viz.} 1\td, \ 6\td, \ 15\td \  and 20\td. The kinematics we have used here is similar to Kheifets {\it et al.} \cite{Kheifets2008} for the plane wave ionization; scattered energy ($E_s$) = 500eV, ejected energy ($E_e$) = 37eV and 74eV and scattering angle ($\theta_s$) = 6\td \ in the co-planar asymmetric geometry.
\begin{center}
	\begin{table}
		\caption{Scaling factors for the twisted electron TDCS of He (1s) atom.} \vskip 5pt
		\resizebox{8cm}{!} 
		{
			\footnotesize
			\addtolength{\tabcolsep}{-8pt}
			\begin{tabular}{ | m{0.05cm} | m{3.0cm}| m{3.0cm} | } 
				\hline
				{\it m} & $E_e$ = 37eV & $E_e$ = 74eV \\ 
				\hline
				1   &  TDCS ($\theta_p$ = 1\td)$\times$ 150 \newline  TDCS ($\theta_p$ = 6\td)$\times$ 10 \newline   TDCS ($\theta_p$ = 15\td)$\times$ 20 \newline   TDCS ($\theta_p$ = 20\td)$\times$ 30 &  TDCS ($\theta_p$ = 1\td)$\times$ 150 \newline  TDCS ($\theta_p$ = 16\td)$\times$ 10 \newline   TDCS ($\theta_p$ = 15\td)$\times$ 20 \newline   TDCS ($\theta_p$ = 20\td)$\times$ 30 \\
				\hline
				2   &   TDCS ($\theta_p$ = 1\td)$\times$ 7$\times10^3$ \newline  TDCS ($\theta_p$ = 6\td)$\times$ 30 \newline   TDCS ($\theta_p$ = 15\td)$\times$ 50 \newline   TDCS ($\theta_p$ = 20\td)$\times$ 80 & TDCS ($\theta_p$ = 1\td)$\times$ 1$\times 10^4$ \newline TDCS ($\theta_p$ = 6\td)$\times$ 50 \newline  TDCS ($\theta_p$ = 15\td)$\times$ 40 \newline  TDCS ($\theta_p$ = 20\td)$\times$ 20 \\
				\hline
				3 & TDCS ($\theta_p$ = 1\td)$\times$ 5$\times10^5$ \newline TDCS ($\theta_p$ = 6\td)$\times$ 50 \newline  TDCS ($\theta_p$ = 15\td)$\times$ 100 \newline  TDCS ($\theta_p$ = 20\td)$\times$ 150 & TDCS ($\theta_p$ = 1\td)$\times$ 1$\times10^6$ \newline TDCS ($\theta_p$ = 6\td)$\times$ 150 \newline  TDCS ($\theta_p$ = 15\td)$\times$ 70 \newline  TDCS ($\theta_p$ = 20\td)$\times$ 40 \\
				\hline
		\end{tabular}}
		\label{table:1}
	\end{table}
\end{center}
\subsection{Angular profiles of the TDCS for plane wave}
Figure \ref{f1}, shows the TDCS as a function of the ejected electron's angle ($\theta_e$) for the plane wave electron impact ionization in the co-planar asymmetric geometry for different noble gas atoms. The momentum transfer and the recoil direction are represented by the arrows ($\theta_{\mathbf{q}}$ and $\theta_{-\mathbf{q}}$) in figure \ref{f1} and subsequent figures.  The left and right columns show the TDCS for $E_e$ = 37eV and 74eV respectively in all the figures. We benchmark the results of our theoretical calculations of the TDCS for the plane wave with the experimental data \cite{Kheifets2008} to validate our theoretical model for the twisted electron impact ionization.\\

\begin{figure*}
	\includegraphics[width=15cm,height=12cm]{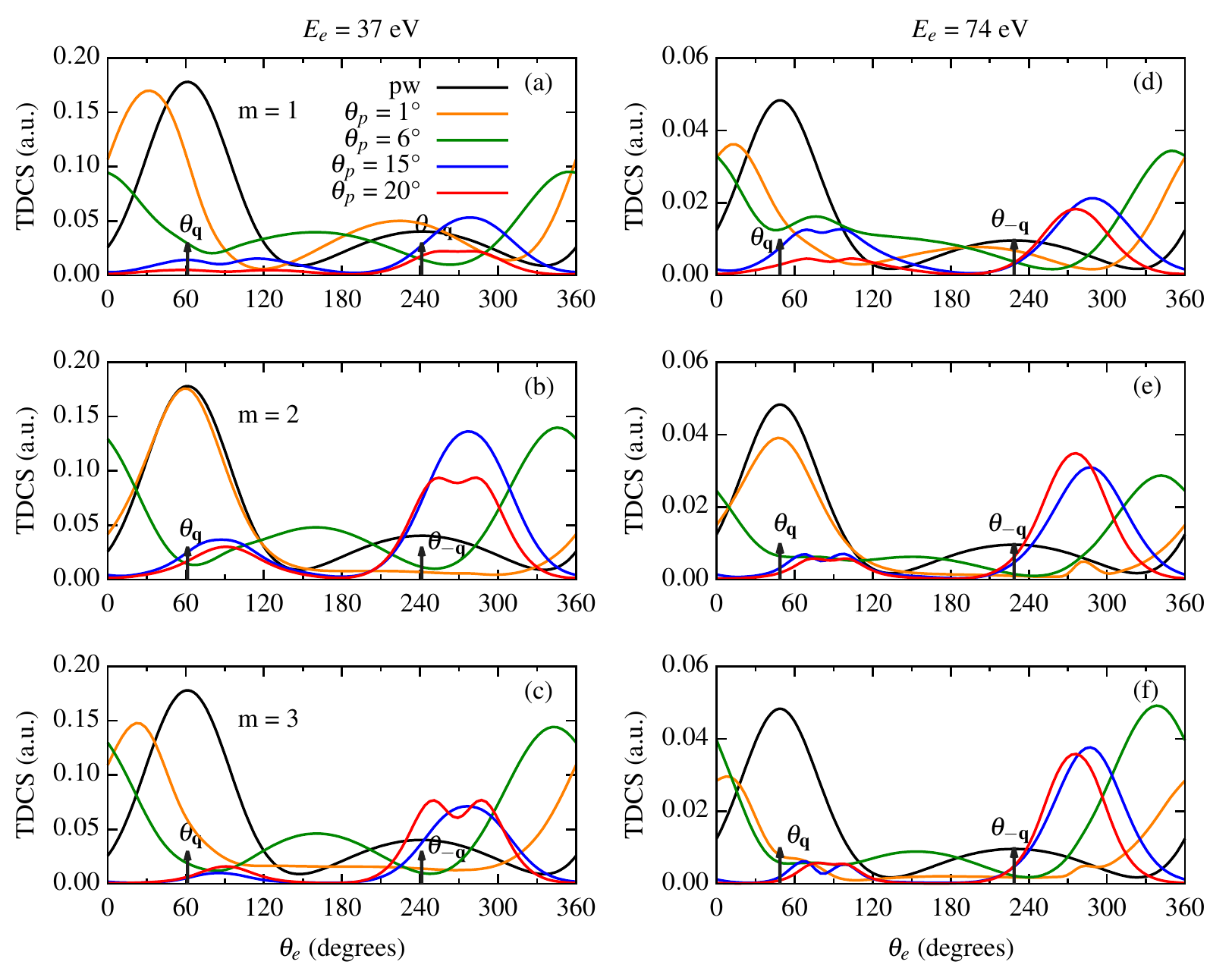}
	\caption{Same as figure \ref{f2} except that the TDCSs are plotted for (e,2e) on Ar (3p) atom.}\label{f5}
\end{figure*}

From figure \ref{f1}, we observe that our theoretical model reproduces the experimental results reasonably well in the binary region for all the cases. The theory, however, underestimates the recoil peak for most of the cases. This may be explained by the fact that our theoretical model uses a 1-Coulomb wavefunction (1CW) for the ejected electron. 
The experimental results are on a relative scale, so we have scaled them to compare the results of our theoretical model in the binary peak region. The plane wave TDCS profile represents a binary peak and a recoil peak along the momentum transfer directions $\theta_{\mathbf{q}}$ and $\theta_{-\mathbf{q}}$ respectively for both the kinematics. For the plane wave ionization, we also observe that the magnitude of the TDCS decreases with an increase in the ejected electron energy. 
\begin{center}
	\begin{table}
		\caption{Scaling factors for the twisted electron TDCS of Ne (2s) atom.} \vskip 5pt
		\resizebox{8cm}{!} 
		{
			\footnotesize
			\addtolength{\tabcolsep}{-8pt}
			\begin{tabular}{ | m{0.05cm} | m{3.0cm}| m{3.0cm} | } 
				\hline 
				{\it m} & $E_e$ = 37eV & $E_e$ = 74eV \\ 
				\hline
				1   & TDCS ($\theta_p$ = 1\td)$\times$ 150 \newline TDCS ($\theta_p$ = 6\td)$\times$ 10 \newline  TDCS ($\theta_p$ = 15\td)$\times$ 20 \newline  TDCS ($\theta_p$ = 20\td)$\times$ 30 & TDCS ($\theta_p$ = 1\td)$\times$ 150 \newline TDCS ($\theta_p$ = 6\td)$\times$ 10 \newline  TDCS ($\theta_p$ = 15\td)$\times$ 20 \newline  TDCS ($\theta_p$ = 20\td)$\times$ 30 \\
				\hline
				2   & TDCS ($\theta_p$ = 1\td)$\times 10^4$ \newline TDCS ($\theta_p$ = 6\td)$\times$ 30 \newline  TDCS ($\theta_p$ = 15\td)$\times$ 40 \newline  TDCS ($\theta_p$ = 20\td)$\times$ 100 & TDCS ($\theta_p$ = 1\td)$\times 15\times 10^3$ \newline TDCS ($\theta_p$ = 6\td)$\times$ 80 \newline  TDCS ($\theta_p$ = 15\td)$\times$ 50 \newline  TDCS ($\theta_p$ = 20\td)$\times$ 50 \\
				\hline
				3 & TDCS ($\theta_p$ = 1\td)$\times$ 1$\times10^6$ \newline TDCS ($\theta_p$ = 6\td)$\times$ 150 \newline  TDCS ($\theta_p$ = 15\td)$\times$ 200 \newline  TDCS ($\theta_p$ = 20\td)$\times$ 300 & TDCS ($\theta_p$ = 1\td)$\times$ 5$\times10^6$ \newline TDCS ($\theta_p$ = 6\td)$\times$ 500 \newline  TDCS ($\theta_p$ = 15\td)$\times$ 150 \newline  TDCS ($\theta_p$ = 20\td)$\times$ 150 \\
				\hline
		\end{tabular}}
		\label{table:2}
	\end{table}
\end{center}

\begin{center}
	\begin{table}
		\caption{Scaling factors for the twisted electron TDCS of Ne (2p) atom.} \vskip 5pt
		\resizebox{8cm}{!} 
		{
			\footnotesize
			\addtolength{\tabcolsep}{-8pt}
			\begin{tabular}{ | m{0.05cm} | m{3.0cm}| m{3.0cm} | }
				\hline
				{\it m} & $E_e$ = 37eV & $E_e$ = 74eV \\ 
				\hline
				1   &  TDCS ($\theta_p$ = 1\td)$\times$ 120 \newline TDCS ($\theta_p$ = 6\td)$\times$ 8 \newline  TDCS ($\theta_p$ = 15\td)$\times$ 20 \newline  TDCS ($\theta_p$ = 20\td)$\times$ 30 & TDCS ($\theta_p$ = 1\td)$\times$ 150 \newline TDCS ($\theta_p$ = 6\td)$\times$ 8 \newline  TDCS ($\theta_p$ = 15\td)$\times$ 20 \newline  TDCS ($\theta_p$ = 20\td)$\times$ 30 \\
				\hline
				2   &  TDCS ($\theta_p$ = 1\td0$\times$ 600 \newline TDCS ($\theta_p$ = 6\td)$\times$ 15 \newline  TDCS ($\theta_p$ = 15\td)$\times$ 60 \newline  TDCS ($\theta_p$ = 20\td)$\times$ 150 & TDCS ($\theta_p$ = 1\td)$\times$ 1500 \newline TDCS ($\theta_p$ = 6\td)$\times$ 25 \newline  TDCS ($\theta_p$ = 15\td)$\times$ 60 \newline  TDCS ($\theta_p$ = 20\td)$\times$ 80 \\
				\hline
				3 & TDCS ($\theta_p$ = 1\td)$\times$ 6$\times10^4$ \newline TDCS ($\theta_p$ = 6\td)$\times$ 40 \newline  TDCS ($\theta_p$ = 15\td)$\times$ 180 \newline  TDCS ($\theta_p$ = 20\td)$\times$ 350 & TDCS ($\theta_p$ = 1\td)$\times$ 1$\times10^5$ \newline TDCS ($\theta_p$ = 6\td)$\times$ 200 \newline  TDCS ($\theta_p$ = 15\td)$\times$ 200 \newline  TDCS ($\theta_p$ = 20\td)$\times$ 300 \\
				\hline
		\end{tabular}}
		\label{table:3}
	\end{table}
\end{center}

\subsection{Angular profiles of the TDCS for the twisted electron wave}
\begin{figure*}[htp]
	\includegraphics[width=16cm,height=15cm]{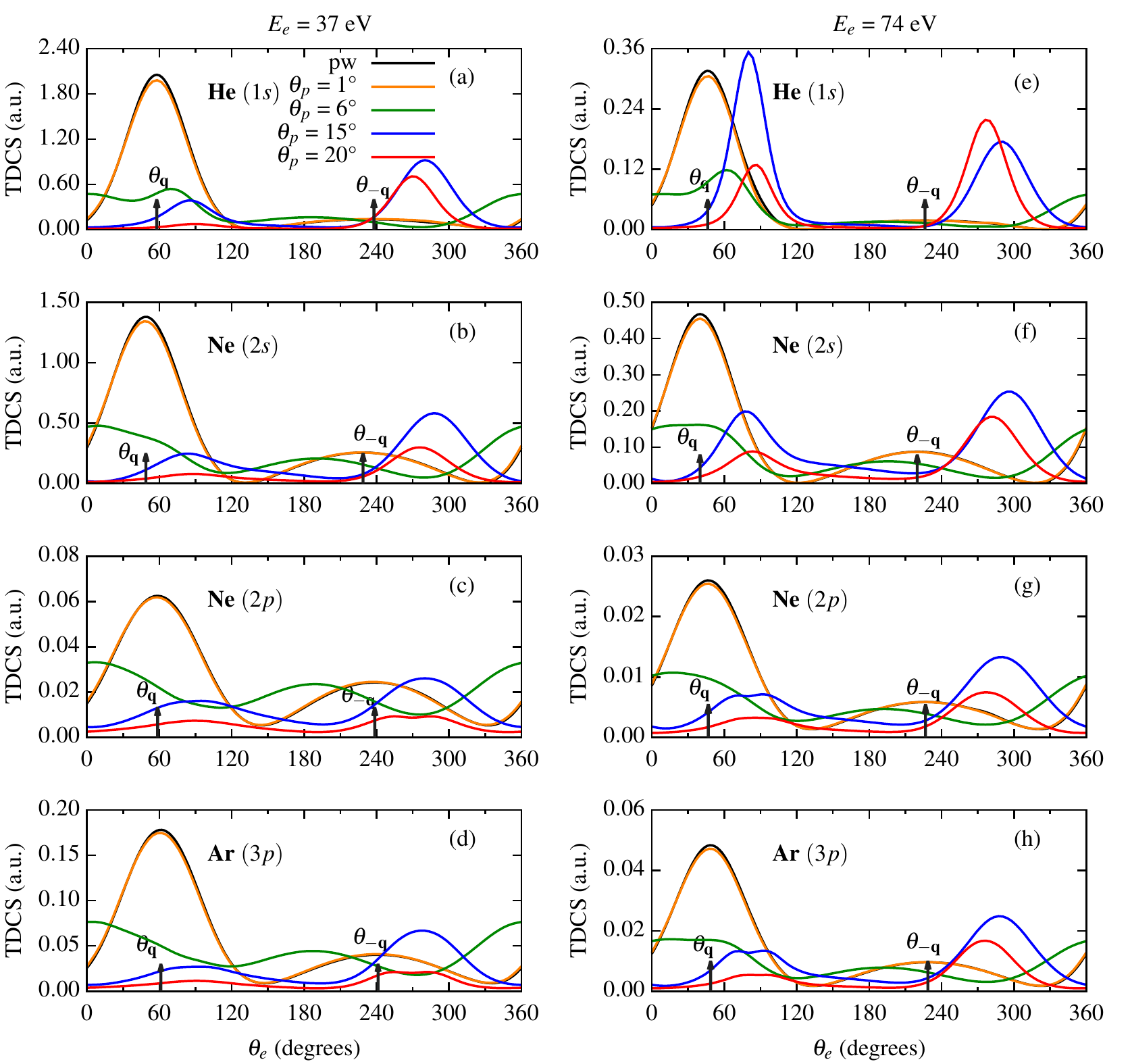}
	\caption{(TDCS)$_{av}$ as a function of the ejected electron angle $\theta_e$ for the plane wave and twisted electron beam for different opening angles (as shown in the frames) respectively. The kinematics used here is same as used in figure \ref{f2}-\ref{f5}. We have scaled up the magnitude of the (TDCS)$_{av}$ by a factor of 5 for Ne (2p) and Ar (3p) for all the values of $\theta_p$.}\label{f6}
\end{figure*}

\vspace{-5ex}
\begin{center}
	\begin{table}[htbp]
		\caption{Scaling factors for the twisted electron TDCS of Ar (3p) atom.} \vskip 5pt
		\resizebox{8cm}{!} 
		{
			\footnotesize
			\addtolength{\tabcolsep}{-8pt}
			\begin{tabular}{ | m{0.05cm} | m{3.0cm}| m{3.0cm} | } 
				\hline
				{\it m} & $E_e$ = 37eV & $E_e$ = 74eV \\ 
				\hline
				1   &  TDCS ($\theta_p$ = 1\td)$\times$ 150 \newline TDCS ($\theta_p$ = 6\td)$\times$ 6 \newline  TDCS ($\theta_p$ = 15\td)$\times$ 20 \newline  TDCS ($\theta_p$ = 20\td)$\times$ 30 & TDCS ($\theta_p$ = 1\td)$\times$ 150 \newline TDCS ($\theta_p$ = 6\td)$\times$ 10 \newline  TDCS ($\theta_p$ = 15\td)$\times$ 20 \newline  TDCS ($\theta_p$ = 20\td)$\times$ 30 \\
				\hline
				2   &  TDCS ($\theta_p$ = 1\td)$\times$ 1200 \newline TDCS ($\theta_p$ = 6\td)$\times$ 20 \newline  TDCS ($\theta_p$ = 15\td)$\times$ 80 \newline  TDCS ($\theta_p$ = 20\td)$\times$ 200 & TDCS ($\theta_p$ = 1\td)$\times$ 1800 \newline TDCS ($\theta_p$ = 6\td)$\times$ 30 \newline  TDCS ($\theta_p$ = 15\td)$\times$ 50 \newline  TDCS ($\theta_p$ = 20\td)$\times$ 90 \\
				\hline
				3 & TDCS ($\theta_p$ = 1\td)$\times$ 1$\times10^5$ \newline TDCS ($\theta_p$ = 6\td)$\times$ 50 \newline  TDCS ($\theta_p$ = 15\td)$\times$ 100 \newline  TDCS ($\theta_p$ = 20\td)$\times$ 400 & TDCS ($\theta_p$ = 1\td)$\times$ 2$\times10^5$ \newline TDCS ($\theta_p$ = 6\td)$\times$ 200 \newline  TDCS ($\theta_p$ = 15\td)$\times$ 150 \newline  TDCS ($\theta_p$ = 20\td)$\times$ 200 \\
				\hline
		\end{tabular}}
		\label{table:4}
	\end{table}
\end{center}
In this section, we present the results of our calculations for TDCS with the twisted electron beam for $E_s$ = 500eV, $E_e$ = 37eV and 74eV and $\theta_s$ = 6\textdegree \  in the co-planar asymmetric geometrical mode for He, Ne (2s and 2p sub-shell) and Ar (3p). Here, we study the effect of twisted electron beam on the TDCS by varying OAM number {\it m} as well as the opening angle $\theta_p$. Figures \ref{f2}-\ref{f5} represent the TDCS for an incident twisted electron beam for the noble gas atoms for  different values of {\it m} and $\theta_p$. In figures \ref{f2}-\ref{f6}, the black, orange, green, blue and red curves represents the TDCS for plane wave (pw), $\theta_p$ = 1\td, 6\td, 15\td \ and 20\td \ respectively. In the figures \ref{f2}-\ref{f5}, frames (a),(d); (b),(e) and (c),(f) depict the TDCS for {\it m} = 1, 2 and 3 respectively for the ejected electron energies 37eV and 74eV.\\

We have used different scaling factors to compare the TDCS for the twisted electron beam with the plane wave TDCS as shown in tables \ref{table:1}-\ref{table:4}.
For all the cases considered here, the magnitude of TDCS is the lowest for the smallest opening angle {\it i.e.} $\theta_p$ = 1\td \ (see scaling factors for $\theta_p$ = 1\td \ in tables \ref{table:1}-\ref{table:4}). The angular profiles for the TDCS at $\theta_p$ = 1\td and for {\it m} = 1 and 3, predict peaks in the binary and recoil regions for He and Ne (2s), and a peak in the binary region only for Ne(2p) and Ar (3p) (see orange curves for {\it m} = 1 and 3 in figures \ref{f2}-\ref{f5}). The peaks, however, are shifted from the momentum transfer direction (see orange curves in figures \ref{f2}-\ref{f5} for {\it m} = 1 and 3 around the arrows). At the same opening angle and {\it m} = 2, we observe the peaks along the momentum transfer directions for He, Ne(2p) and Ar (see orange curves in figures \ref{f2}, \ref{f4}, \ref{f5} for {\it m} = 2 around the arrows). However, for Ne (2s) at $\theta_p$ = 1\td \ and {\it m} = 2, we observe peaks in the forward and backward regions (see orange curves in figure \ref{f3} (b) and (e)).\\

When the opening angle is same as the scattering angle {\it i.e.}, $\theta_p$ = $\theta_s$, we observe peaks in the forward and backward direction for most of the cases (see green curves in the figures \ref{f2}-\ref{f5} around $\theta_e$ = 0\td(or 360\td) and 180\td). However, for He and Ne (2s), we observe peaks in the forward and perpendicular directions for {\it m} = 1 only (see green curves in the figures \ref{f2} (a) and (d) and \ref{f3}  (a) and (d) around $\theta_e$ = 0\td(or 360\td) and 90\td). For all the cases, a minimum is observed around the plane wave momentum transfer direction. Similar observation has also been accounted in our previous communication for the water molecule \cite{Dhankhar2021} and by Harris {\it et al.} for the hydrogen atom \cite{Harris2019}. \\
On further increasing the opening angle of the incident beam to 15\td, some interesting results are observed. The magnitude of the TDCS decreases as $\theta_p$ is increased from 6\td. For the {\it s}-orbitals (He and Ne (2s)), the angular profile of the TDCS at $\theta_p$ = 15\td \ represents a two peak structure around the perpendicular directions with the dominant peak close to $\theta_e$ = 270\td \ and a shallow peak around $\theta_e$ = 90\td \ (see blue curves in figures \ref{f2} and \ref{f3}), except for $E_e$ = 74eV and {\it m} = 1. For $E_e$ = 74eV and {\it m} = 1, the dominant peak is around 90\td \ (see blue curve in figure \ref{f2}(d) and \ref{f3}(d) around $\theta_e$ = 90\td \ and 270\td). For Ne (2p) and Ar (3p) at $\theta_p$ = 15\td, the peak around $\theta_e$ = 90\td \ splits and we observe a double-peak structure which is a characteristic of the {\it p}-dominant atomic orbitals. The double-peak structure in the TDCS is observed only for {\it m} = 1 for $E_e$ = 37eV and for all values of {\it m} in the TDCS for $E_e$ = 74eV for Ne (2p) and Ar (3p) (see blue curves in figures \ref{f4} and \ref{f5}).\\  

For an opening angle $\theta_p$ = 20\td, the angular profiles of the ionization cross-sections is somewhat similar to those observed for $\theta_p$ = 15\td. We observe peaks approximately around $\theta_e$ = 90\td \ and 270\td\ as observed earlier for He and Ne (2s) (see red curves in figures \ref{f2} and \ref{f3}). For Ne (2p) and Ar (3p), at lower energy of the ejected electron {\it i.e.} $E_e$ = 37eV, we observe a shallow peak and a double-peak structure (similar to the one observed at $\theta_p$ = 15\td \  for the same atoms) around $\theta_e$ =  90\td \ and 270\td \ respectively for most of the cases.
However, for $E_e$ = 74eV and {\it m} = 1, we observe a dominant peak and a shallow double-peak structure around $\theta_e$ = 270\td \ and  90\td \ respectively.
For the same ejected electron energy and {\it m} = 2 and 3, we observe a dominant peak and a shallow peak in the perpendicular directions (see red curves in figures \ref{f4}-\ref{f5} around $\theta_e$ = 90\td \ and 270\td).  Due to the additional transverse momentum, we observe peak splitting for the twisted electron TDCS at higher opening angles only, {\it i.e.}, for $\theta_p$ = 15\td \ and 20\td \ for the Ne (2p) and Ar (3p).
The present calculations of the TDCS for twisted electron wave on the noble gas atoms clearly show the dependence of TDCS on  both {\it m} and $\theta_p$.
\vspace{-3ex}
\subsection{Angular profiles for the (TDCS)$_{av}$ for  macroscopic atomic targets}

In figure \ref{f6}, we present the results of our calculations for the TDCS averaged over the impact parameter \textbf{b}, (TDCS)$_{av}$, as a function of the ejected electron angle $\theta_e$ for He, Ne (2s and 2p shells) and Ar (3p) noble gases in the co-planar asymmetric geometry. As shown by the equation \ref{eq19}, the (TDCS)$_{av}$ depends only on the opening angle $\theta_p$ of the incident twisted electron beam. We keep the same kinematics as used in the earlier figures. We compare the results of the averaged cross-section ((TDCS)$_{av}$) with the plane wave TDCS. We present the angular profiles for $\theta_p$ = 1\td, 6\td, 15\td \  and 20\td. For a better comparison of the (TDCS)$_{av}$ with the plane wave TDCS, we have scaled the (TDCS)$_{av}$ for $\theta_p$ = 15\td \ and 20\td \ by a factor of 5. We have not used any scaling factors for $\theta_p$ = 1\td \ and 6\td.  From figure \ref{f6}, we observe that for the smallest opening angle, {\it i.e.} $\theta_p$ = 1\td, the angular profile of the (TDCS)$_{av}$ is similar to that of the plane wave TDCS for all the atoms (see orange curves in the figure \ref{f6}).  For $\theta_p$ = $\theta_s$ = 6\td, we observe peaks in the forward and backward direction with a small peak around the plane wave momentum transfer direction for He only (see green curves around the arrows in figure \ref{f6} (a) and (e)). For higher opening angles, like 15\td \ and 20\td, we observe that the angular profiles represent a two peak structure with peaks around the perpendicular directions ($\theta_e$ = 90\td \ and 270\td). For both the opening angles, the dominant peak is approximately around $\theta_e$ = 270\td \ and the shallow peak is located close to $\theta_e$ = 90\td \ except for He, wherein for $E_e$ 74eV and $\theta_p$ = 15\td \ the dominant and shallow peak is located around $\theta_e$ = 90\td \ and 270\td \ which is in contrast to all the other cases for $\theta_p$ = 15\td. We also observe that the magnitude of (TDCS)$_{av}$ decreases with an opening angle.  
It can be seen from equations \ref{eq2} and \ref{eq12} that the plane wave TDCS is inversely proportional to the quad of momentum transfer magnitude. Thus, in computing the (TDCS)$_{av}$, the plane wave TDCS from azimuthal angles with smaller momentum transfer magnitude will dominate the average, thus resulting in a shift of binary peak to the recoil direction for higher $\theta_p$ (for more details see \cite{Plumadore2020}).


\section{Conclusion}\label{sec4}
In this paper, we have presented the theoretical study of the triple differential cross-sections (TDCS) for an (e,2e) process on different noble gas atoms (He, Ne (2s and 2p), and Ar (3p)) by the twisted electron beam. We studied the angular distributions of the TDCS for the co-planar asymmetric geometry in the first Born approximation for both the plane and twisted electron beam. We have studied the effect of the OAM number, {\it m} and the opening angle $\theta_p$ of the incident twisted electron beam on the TDCS. We have benchmarked our theoretical results with the experimental data for the plane wave electron beam.\\
For the (e,2e) ionization of the noble gas atoms by twisted electron impact, the angular profile of the TDCS is altered from the plane wave case. We also observe that the magnitude of the TDCS for twisted electron wave impact is smaller. For the plane waves, the angular profile of the TDCS peaks around $\theta_\mathbf{q}$ (binary peak) and $\theta_{-\mathbf{q}}$ (recoil peak). However, for the twisted electron beam, due to the additional transverse momentum and the phase (OAM ({\it m})) dependence of the incident twisted electron beam, the angular profiles of the TDCS for twisted electron case the peaks are shift from the momentum transfer directions for a smaller opening angle. A minimum occurs around the plane wave momentum transfer direction when $\theta_p$ = $\theta_s$. For higher opening angles, like 15\td \ and 20\td, we observe peaks in the perpendicular directions. At these opening angles, we also observe that the peaks split into double-peak structures for {\it p}-type orbitals (Ne (2p) and Ar (3p)). 
In addition, we discuss the (TDCS)$_{av}$ (averaged over the impact parameter \textbf{b}) as a function of the opening angle $\theta_p$ of the twisted electron beam. For a macroscopic target, the angular profiles of (TDCS)$_{av}$ significantly depend on the opening angle ($\theta_p$) of the twisted electron beam. At smaller opening angle, like $\theta_p$ = 1\td, the (TDCS)$_{av}$ is similar to the plane wave TDCS. However, for higher opening angles {\it i.e.} 15\td \ and 20\td \ the peaks shift to perpendicular directions. The results of (TDCS)$_{av}$ show the dependence of momentum transfer and thus the averaged cross-section on the opening angle of the incident twisted electron beam.

Our present communication attempts to investigate the (e,2e) process on the noble gas atoms to unravel the effects of the twisted electron's different parameters on the angular profile of the TDCS. We have used the 1CW wave function in our theoretical model to study the TDCS. The present study can also be extended for other complex theoretical models, like DWBA, CCC, 2C, etc.. We are working on this aspect.	


%

\end{document}